\documentclass[10pt,twoside]{article}

\usepackage{graphicx}
\usepackage[T1]{fontenc}
\usepackage{physics}
\usepackage{amsmath}
\usepackage{amssymb}
\usepackage{xcolor}
\usepackage{array}
\usepackage{verbatim} 
\usepackage{latexsym}
\usepackage[english]{babel}
\usepackage[pdfa]{hyperref}
\def\volnum{Manuscrit}
\usepackage{aflbcours}
\usepackage{enumitem}
\setitemize{noitemsep,topsep=0pt,parsep=0pt,partopsep=0pt}

\usepackage{etoolbox}
\AtBeginEnvironment{quote}{\vspace{-0.5\baselineskip}}
\AtEndEnvironment{quote}{\vspace{-0.5\baselineskip}}

\pdebut{1}
\def\pacs#1{\LP P.A.C.S.: #1}
\title{The Solvay Councils, de Broglie’s brothers, 
and the development of wave-particle duality.}
\titleshort{De Broglie-Solvay}

\author{Alessio Rocci* -- Franklin J. Lambert**}
\authorshort{A. Rocci, F.J. Lambert}
\address{(*) Theoretische Natuurkunde (VUB) and the International Solvay Institutes; Applied Physics research group (APHY)\\
(*) Vrije Universiteit Brussel (VUB) and the International Solvay Institutes,\\

(*) alessio.rocci@vub.be -- (**) franklin.james.lambert@gmail.com}

\begin{document}

\maketitle

\begin{abstract}
The meeting patronized by Ernest Solvay in 1911, the first Solvay Council, marked the beginning of what can be called the \textit{first quantum revolution}. Maurice de Broglie was one of the secretaries of the Council. He participated in the two following conferences and contributed to the third Council with his work. Louis de Broglie collaborated and discussed with his elder brother. Departing from Maurice's achievements, Louis developed his approach to the wave-particle duality of light and extended it to other particles. He presented it at the fifth meeting in 1927. This paper investigates the connection between the two scientists and the Solvay Councils, arguing that this link was essential in developing the first quantum revolution, a process that ended, according to Werner Heisenberg, at the fifth Solvay Physics Council. 
\end{abstract}
\pacs{03.65.Bz; 04.20.-q}

\section*{Introduction}
Max Planck’s announcement of his law addressing blackbody radiation data at the beginning of the twentieth century marked the introduction of the quantum of action in physics. However, its necessity emerged only in the following ten years. According to the historian Dieter Hoffman\footnote{We are providing an English translation of Hoffman’s German text.}: “Quantum theory was by no means part of the universally secured body of physical knowledge at that time, […] its general recognition had only begun with the Solvay Congress of 1911.” (\cite{Hoffmann}, p. 17) 
The first Solvay conference, henceforth Solvay I, and the following meetings offered a unique framework for addressing experimental and theoretical questions related to the problem of quanta. This paper focuses on the sixteen years separating Solvay I (1911) and the fifth meeting, henceforth Solvay V, held in Brussels in 1927. The Solvay Councils between these two meetings will be referred to as Solvay II (1913), Solvay III (1921), and Solvay IV (1924).\footnote{In the following, we will consider the Solvay Physics Council only, avoiding confusion with the Chemistry conferences that started in 1922.} All the scientists that contributed to the transition from the old quantum theory to new quantum mechanics gathered at Solvay V (1927). Louis de Broglie’s introduction of wave-particle duality (1924), the discovery of matrix-mechanics by Werner Heisenberg, Max Born, and Pascual Jordan (1925), and Erwin Schr\"odinger’s wave mechanics (1926) merged into quantum mechanics after Solvay V.

The proof of equivalence between the two theoretical schemes by Schr\"odinger and Heisenberg’s paper with his uncertainty relations were both published before another meeting in Como. However, as Heisenberg emphasized in his interview with Thomas Kuhn, the Como conference had not been nearly so important as the Brussels meeting \cite{Kuhn-1}: 
\begin{quote}
    In Como, there were some very informal talks, and I think I had to give a lecture. Schr\"odinger gave a lecture. Certainly, the Como meeting did a lot to spread the knowledge of this new scheme to many physicists. So it was the first time that a large group of physicists would listen to that kind of thing and hear that now there is a development going on that seems to be in a state of clarification. The Brussels meeting was where the details were really discussed and where one had to fight.
\end{quote}

Heisenberg referred to the so-called Bohr-Einstein debate. Even if the disagreement between Niels Bohr and Albert Einstein continued during the following Solvay Council (1930), Heisenberg underlined that the meeting of 1927 marked the end of the process for a specific group of physicists \cite{Kuhn-1}: “This group, Bohr, Ehrenfest, Pauli, and myself, felt that they were on the right track, that they had a good concept and no lack of clarity anymore. So we were all quite happy and felt that the game was now won.” To describe this situation, Heisenberg often emphasized that the Solvay V certified the \textit{completion} of quantum mechanics. Heisenberg was referring to the fact that a cycle was concluded for the proponents of the new mechanics. However, as it is well known, some questions were not addressed at Solvay V. One was the concept of \textit{entanglement}, which became one of the pillars of the \textit{second} quantum revolution, which started in the mid-1950s.

The sixteen years necessary to achieve the conceptual leap from classical to quantum physics (1911-1927) coincide with the period during which the Scientific Committee of the \textit{International Solvay Institute of Physics} (ISIP) was placed under the enlightened chairmanship of Hendrik A. Lorentz. This striking coincidence aroused our interest in the role played by \textit{Ernest Solvay’s Science Project} during this golden age. We call \textit{Ernest Solvay’s Science Project} the road map suggested by Ernest Solvay and implemented by Lorentz, which allowed the ISIP to grant subsidies\footnote{Another noteworthy action of ISIP was the bursary program, which allowed young Belgian scholars to perfect their skills in foreign laboratories. This local aspect of ISIP’s efforts contributed to developing physics in Belgium. For a description of these aspects of ISIP’s program, see \cite{Lambert-Berends}.} to experimental physicists from all nations and to promote the exchange of ideas between the most competent researchers through the regular organization of Physics Councils by an International Scientific Committee (ISC).

Maurice de Broglie and his younger brother Louis participated in this adventure from the very beginning. However, they only partially contributed to Solvay’s project. Indeed, during Lorentz’s presidency, they were never called to participate in the ISC, and due to the wealthy financial situation of de Broglie’s family, they never profited from ISIP’s subsidy program. Despite this, they contributed to the discussions of the Solvay meetings. This paper aims to describe the connection between the Councils and the work of the two brothers\footnote{To distinguish between the two de Broglie brothers, we will use their names: Maurice and Louis.}. We argue that Maurice’s work for the Solvay meetings proved crucial for transitioning from the old quantum theory to quantum mechanics. As it is well known, Maurice influenced his brother’s scientific work. We present an in-depth analysis of how Maurice's work affected Louis' viewpoint. We investigate the link between the proceedings of the Solvay conferences, Maurice's experimental work, and Louis' development of his wave-particle approach.

The paper is organized as follows. In section \ref{Solvay-I}, we present Solvay I (1911) and describe some elements that will emerge in Louis' work. In section \ref{X-section}, we briefly explain how the discoveries of Max von Laue and the Bragg family on X-rays affected Solvay II (1913). Then, we describe their impact on Maurice's work and the preparation of his work for Solvay III (1921). Section \ref{legacy} analyzes the legacy of the third Council and Louis' early scientific production. We describe some hidden connections between the de Broglie brothers and the interpretation of the Compton effect. In section \ref{Solvay4}, we argue that Solvay IV (1924) was the first moment to spread Louis' innovative ideas. We conclude our analysis with section \ref{Solvay5}, where we quote some less-known Louis' recollections about his encounter with Einstein at Solvay V (1927). The paper finally ends with a brief epilogue, which explains why Louis decided to stop improving his approach at that time.

\section{Solvay I: The theory of radiation and quanta (1911)}\label{Solvay-I}
On June 15, 1911, twenty prominent physicists received a letter entitled: \textit{Invitation to an International Scientific Council to elucidate topical issues of molecular kinetic theory}. The invitation was confidential. It was signed “E. Solvay.” Today, the first Solvay Council, hereafter Solvay I, held in Brussels from October 30 to November 3, 1911, is cited as history’s first international conference on quanta. Indeed, it convened initially to discuss molecular kinetic theory, but the reports and discussions focused on problems related to Einstein’s introduction of the light quantum hypothesis. Curiously, this “private conference” in physics was convened by an industrialist and took place in a country where no one cared about the molecular implications of Max Planck’s quantum hypothesis. In fact, the idea of an \textit{International Scientific Council} was launched in 1910 by Walther Nernst, the Institute of Physical Chemistry director at the University of Berlin\footnote{Nernst’s motives for planning this unprecedented “Council in Physics” are reported in detail in (\cite{Lambert-Berends}, p. 30-40).}. 

The timing of the plan proved crucial. Ten years had passed since Planck discovered the elementary quantum of action. Solvay got involved in organizing a Physics Council when measurements of specific heats, carried out by Nernst, showed that Planck’s hypothesis could be made fruitful in a domain that differed from the one in which it had been introduced. Hence, by publicizing the successful exportation of the quantum hypothesis from the field of radiation to the field of matter, the Council appeared to Nernst as a means of fostering its acceptance. It also seemed a means to bring the quanta to the attention of a large community of physicists and chemists. Nernst’s idea, conceived for personal reasons, proved most effective. Inspired by the Council’s discussions, Henri Poincaré wasted no time demonstrating the need to accept the notion of energy quanta \cite{Poincare}.  

\subsection{Louis and Solvay I}\label{Louis-Solvay1}
The scientific life of Maurice and Louis de Broglie was affected by the Solvay Physics Councils from the beginning. According to the historian Bruce Wheaton, Henri Poincaré invited the elder brother Maurice to be one of the scientific secretaries of Solvay I (\cite{Wheaton-article}, p. 40). At the time, the two brothers had already lost their father: Maurice was 36 years old, and his younger brother Louis just turned 19. As Duke Philippe Maurice de Broglie confirmed\footnote{We acknowledge Duke Philippe Maurice for sharing this information with us during his visit to the Solvay Archive in Brussels.}, his ancestor Maurice took Louis as his secretary in Brussels. Louis had already completed his history license, and his reading of Poincaré’s books, stimulated by his brother, had convinced him to study physics. Regarding the role of his brother, in 1931, Louis wrote (\cite{LdB-1931}, p. 15): 
\begin{quote}
    My natural inclination was toward mechanics, theoretical physics, and the philosophy of science. But under the guidance of my brother Maurice and living in the atmosphere of the laboratory where he pursued his experimental research, I was from that time on accustomed to not losing sight of the fact that the theoretical constructs of science are only of value if they are always based on facts and remain in contact with them.
\end{quote}
Moreover, according to later recollections, ever since Louis turned to the study of theoretical physics at the age of nineteen, he has been “an ardent admirer of both the person and the work of Albert Einstein.” (\cite{LdB-1976}; p.14).

Einstein was one of the main protagonists of Solvay I. He reported on the problem of specific heats and expanded his review to embrace the idea of light quanta. Once back in Paris, the young Louis had the opportunity to study the reports and read in-depth the discussions at the Council very early because Maurice was charged with their publication with Paul Langevin \cite{Minutes}. When Louis had the Solvay minutes in his hands, he reacted as follows (\cite{LdB-Penseur}; p. 458):
\begin{quote}
    With the enthusiasm of my age, I had been thrilled by the interest in the problems studied, and I had promised myself to devote all my efforts to understanding \textit{the true nature of the mysterious quanta} […] At the time, I had already perceived the importance […] of \textit{Analytical Mechanics} [emphasis added]. 
\end{quote}
The above-emphasized aspects will become the focus of Louis de Broglie’s Ph.D. thesis.   

Louis emphasized the following points in a report on his work dated 1931. “I read the reports [of Solvay I] with passionate interest. Reading Mr. Planck’s report introduced me to the mysteries of quantum theory, the development of which has dominated the entire evolution of theoretical physics over the last thirty years” (\cite{LdB-1931}, p. 14). Louis stressed the role of Einstein as follows (\cite{LdB-1931}, p.15): 
\begin{quote}
  The discovery of the photoelectric effect, in particular, had led Mr. Einstein to assume that all frequency radiation is made up of $h\nu$-valued corpuscles, a new mysterious intervention of the constant h. Revived from the Greeks and Newton, this corpuscular theory of light seemed irreconcilable compared to the wave theory, which was firmly established based on diffraction and interference experiments. This fact heralded a formidable crisis in theoretical physics.  
\end{quote}
His brother Maurice similarly emphasized: “The role played by the general principles of Analytical Mechanics had struck him.” (\cite{MdB-Penseur}; p. 425). Regarding the role of Analytical Mechanics, Louis recollected (\cite{LdB-1931}; p. 15):
\begin{quote}
    Looking through the minutes of the Solvay Council, I was struck by the fact that the general theories of Analytical Mechanics played an essential role in developing the quantum doctrine. Although the constant $h$ is utterly foreign to classical Analytical Mechanics, the latter seems to form a ready-made mold for it. Entirely determined to devote myself to studying the quantum mystery, I made Analytical Mechanics and its relationship with quanta the focus of my reflections. 
\end{quote}

In the following section, we briefly summarize the content of Solvay I with an emphasis on the reports presented by Lorentz, Planck, and Einstein. The first stressed how the approach of Analytical Mechanics should have been modified. Planck presented the puzzles connected with the introduction of light quanta in a very pedagogical way. Einstein introduced his brand-new findings to support the light-quanta hypothesis.

\subsection{The proceedings of Solvay I}\label{proceedings-Solvay1}
At the opening of Solvay I, Lorentz emphasized the depth of the crisis: “We have the feeling that we are at a dead end, as the old theories are showing themselves to be powerless to pierce the darkness which surrounds us at all sides” \cite{proceedings-Solvay1}. The presentation of eleven reports preceded the deliberations of the Council\footnote{The discussion of every report took place during its lecturing \cite{Minutes}. However, in the official proceedings, the discussions have been placed at the end of the corresponding report.}:
\begin{itemize}
    \item[1] \textit{Application of the energy equipartition theorem to radiation} (Lorentz); 
    \item[2] \textit{On Clausius, Maxwell, and Boltzmann's kinetic theory of specific heat} (James Jeans);
    \item[3] \textit{Experimental verification of Planck’s formula in the region of high frequencies} (Emil Warburg);
    \item[4] \textit{Planck’s radiation formula's verification in the long wave region} (Heinrich Rubens);
    \item[5] \textit{The blackbody radiation law and the hypothesis of elementary action quantities} (Planck); 
    \item[6] \textit{Kinetic theory and the experimental properties of ideal gases} (Martin Knudsen);
    \item[7] \textit{Evidence of molecular reality} (Jean Baptiste  Perrin); 
    \item[8] \textit{Application of the theory of quanta to a number of physicochemical and chemical problems} (Nernst);
    \item[9] \textit{Application of the “element of action” theory to non-periodical molecular phenomena}  (Arnold Sommerfeld);
    \item[10] \textit{Kinetic theory of magnetism, the magnetons} (Langevin);
    \item[11] \textit{The current state of the specific heat problem} (Einstein).
\end{itemize}
Most reports, notably those by Planck, Nernst, Sommerfeld, and Einstein, led to vivid discussions. Lorentz translated questions and answers. He summarized the debates at the end of each session. Maurice took note of the interventions made in French. He collected handwritten notes from members who spoke in German or English \cite{Minutes}. It was agreed that the Council’s Proceedings would be published in French out of respect for Mr. Solvay. Langevin was asked to take care of the translations.

The volume, edited by Langevin and Maurice, appeared under the title \textit{The theory of radiation and quanta}. The contrast with the Council’s initial goal -- to elucidate topical issues of molecular kinetic theory -- reflects a clear shift of focus in the debates\footnote{The French titles read: \textit{La théorie du rayonnement et les quanta} and \textit{Conseil pour élucider quelques questions d’actualité des théories cinétiques et moléculaires}.}. The more fundamental radiation problems slowly overshadowed questions regarding kinetic and molecular theory.

Maurice analyzed the meeting in a booklet published in 1951. “Lorentz insisted on the difficulties connected with classical theories […] [He] discussed Planck’s ideas and showed their necessity despite their strangeness” (\cite{MdB-1951}, p. 22). The importance of Analytical Mechanics, the Lagrangian methods, and Hamilton’s equations emerged in Lorentz’s report. The discussion addressed the applicability of Hamilton’s equations in the context of new quantum phenomena\footnote{The applicability was questioned in connection with the equipartition and Liouville’s theorems for canonical ensembles.}. Marcel Brillouin, one of the participants, underlined the same point in his annotated version of the Council’s reports\footnote{Document 11127; PLS-ESPCI \url{https://bibnum.explore.psl.eu/s/psl/ark:/18469/1z19r}.}. Louis would study physics under Brillouin in Paris after the Great War.

Brillouin highlighted another comment in Lorentz’s report: “We should invent a mechanism enabling the energy transfer in the form of rapid vibrations from the ether to matter excluding the reverse” (\cite{proceedings-Solvay1}, p. 13). This statement clearly expresses the asymmetry in the process described from Lorentz’s perspective at the time. Maurice would decisively assert the symmetry in the energy exchanges at Solvay III (1921).

Planck exposed an alternative to his initial theory, the so-called \textit{emission theory}, in which he had recently developed an attempt to reconcile his radiation law with classical electrodynamics and presented alternatives to his theory. In his booklet on the Solvay Councils, Maurice emphasized Planck’s hesitancy in accepting light quanta when presenting Einstein’s ideas. Brillouin underlined the same statement in his annotated reports. Planck’s comment is as follows (\cite{proceedings-Solvay1}, p. 93):
\begin{quote}
    Einstein adopted the first point of view in his hypothesis of light quanta. Stark followed it. According to this hypothesis, the energy of a light ray of frequency $\nu$ is not distributed continuously in space, and it propagates in a straight line by determined quanta of magnitude $h\nu$ in the same way as light particles in Newton’s theory of emission. An important confirmation of this hypothesis is that the velocity of secondary cathode rays produced by Roentgen rays is independent of the intensity of these rays. J.J. Thomson’s study of photoelectric phenomena led him to a similar conception; he believed he could only explain the small number of electrons emitted and the independence between their speed and the intensity of the incident light by admitting, instead of a uniform distribution of energy across the front of the light waves, local accumulations of this. Needless to say, such hypotheses were irreconcilable with Maxwell’s equations and all the electromagnetic theories of light proposed to date. 
\end{quote}    
Maurice underlined the association between the names of Einstein and J. J. Thomson in his book entitled \textit{Les Rayons X} (1922). Langevin would do the same in evaluating Louis' Ph.D. dissertation. The historian Wheaton highlighted Langevin's note without offering a possible explanation. The above quotation shows that Solvay I Planck’s report was the common denominator. Brillouin also reacted to Planck's remark in his annotated minutes of Solvay I by sketching non-uniformly distributed wavefronts.

In the last report to the Council, Einstein finally analyzed the specific heat problem and discussed his recent findings on thermal radiation and energy grains. Einstein’s conclusions about the issue of specific heats did not convey the positive message that Nernst expected. Instead of emphasizing the success of Planck’s hypothesis in the molecular domain, Einstein noted that in its current form, quantum theory could not explain the behavior of thermal conductivity of matter in the vicinity of the absolute zero of temperature. Extending his report beyond the problem of specific heat, Einstein addressed the fundamental issue of thermal radiation. Regarding Planck’s law, he made this discouraging remark: “I had no other goal here than to show how fundamental are the difficulties in which the radiation formula leads us, even if we consider it as a simple experimental fact” (\cite{proceedings-Solvay1}, p. 407). 

Regarding Planck’s theory, he drew a dark picture of the challenge ahead  (\cite{proceedings-Solvay1}, p. 407):
\begin{quote}
    These discontinuities, which make it so difficult to accept Planck’s theory, seem to exist in nature. The difficulties that a satisfactory theory of these fundamental phenomena must overcome seem, at this moment, insurmountable. Why does an electron in a metal struck by R\"ontgen rays take the high kinetic energy observed in the secondary cathode rays? All the metal is in the field of the R\"ontgen rays; why is it only a small part of the electrons that pick up the speed of these cathode rays? How come the energy is absorbed in an extraordinary few points? How do these few points differ from others? We have no answer to these questions and to so many others.
\end{quote}

Einstein introduced his theoretical consideration on the hypothesis of quanta as follows. “Now we come to the all-important, but unfortunately still unresolved question: how can Mechanics be modified to align it with the law of radiation and the thermal properties of matter?” (\cite{proceedings-Solvay1}, p. 407). Again, Einstein’s central point was the application of statistical mechanics and his blackbody fluctuation formula. He showed that two distinct contributions are needed to compute the energy fluctuations described by Planck’s formula: one represented the contribution due to the undulatory character of electromagnetic radiation -- like in Maxwell’s theory -- while the other was due to the particle properties of light. Einstein’s formula was one of Louis' starting points when he turned his attention to the blackbody problem after the Great War, and he always quoted the proceedings of Solvay I as his source.

Brillouin was impressed by the same point when studying Einstein’s report. In his annotated version of the Solvay minutes, he underlined the fluctuation formula and its meaning and Einstein’s reference to light quanta as energy plots. By accepting the hypothesis that light energy is localized, Brillouin supported the idea of quanta as particles and deprived Maxwell’s waves of part of their reality character. Brillouin and Louis followed this path to reconcile light’s wave and particle aspects. The following remark made by Brillouin during the final discussion reflects the general feeling at the end of the conference (\cite{proceedings-Solvay1}, p. 451):
\begin{quote}
    It now seems quite certain that it will be necessary to introduce into our physical and chemical concepts a discontinuity, an element varying by leaps, of which we had no idea a few years ago. How should it be introduced? This is what I see less well. [...] I don’t know. The uncertainty in which we find ourselves about the form and the extent of the transformation that must be operated - evolution or a complete overhaul - is a powerful stimulus. This concern will certainly haunt us for many weeks, and each of us will passionately focus on solving the difficulties that our discussions have shown to be inevitable and important in so many fields of physics and chemistry. 
\end{quote}

\section{X-rays: to the heart of atoms}\label{X-section}
Louis was too young when he first encountered the ideas discussed at Solvay I. He recollected (\cite{LdB-1976} p. 14): 
\begin{quote}
    When, with increased maturity, I was able to return to my studies after the long interruption caused by World War I, it was once again the ideas of Einstein, this time on the wave-particle dualism in light, that guided me in my attempts to extend this dualism to matter.
\end{quote}

Louis' approach is deeply rooted in the experimental work of his elder brother. Maurice started his work on X-rays soon after von Laue and William H. Bragg’s discovery (1912). The reflection and diffraction of X-rays by crystals were discussed at Solvay II (1913). Again, Maurice was one of the three secretaries with Frederick Alexander Lindemann. Both had a more active role this time and presented some experimental results. Soon after Solvay II, Maurice devised an extremely elegant experimental technique -- the \textit{rotating crystal} method -- which made it possible, by dropping a beam of X-rays onto a slowly rotating crystal, to photographically record the diffraction spectra and thus obtain the spectral analysis of this radiation, as well as essential information on the crystal structure. Right from the start of his research, he discovered several fundamental laws governing the emission and absorption of X-rays. He confirmed Henry Moseley’s law by extending it to many elements. Even if Bragg is usually recognized as the father of X-ray spectroscopy, according to Jean-Jacques Trillat, Maurice deserves a similar title. Indeed, Trillat underlined that thanks to his progress, Maurice clarified the structure of atoms, concluding as follows: “Maurice de Broglie was the father of X-ray spectrography.” (\cite{Trillat-1960}; p. 239) All these elements clarify why Rutherford indicated Maurice’s name to be one of the rapporteurs at Solvay III (1921).

In 1919, Langevin lectured on the theory of quanta at Collége de France. The annotations Louis took and described by the historian Chieko Kojima can be reconsidered from our perspective. Louis' annotations are focused on Bohr’s theory, Sommerfeld’s quantization rules, the Jacobi function, and the Hamilton-Jacobi theory. These arguments are coherent with the interests of the young de Broglie stimulated by the Solvay meetings. Sommerfeld’s ideas were presented at Solvay I (1911), and the Hamilton-Jacobi theory is one of the pillars of analytical mechanics. There is no track of Bohr’s theory in the proceedings of Solvay II. However, it was probably discussed in private. This subject soon became the center of Maurice’s research program. The very first communication published in 1920 by Louis was devoted to the experimental verification of Bohr’s model.

Chieko Kojima noticed that in his lectures, Langevin did not mention Einstein’s latest research into the nature of radiation. However, Louis would quote Einstein’s 1917 note on radiation in his second published communication, showing to be updated on Einstein’s work. This fact is unsurprising because, as Louis recollected, Einstein’s figure always fascinated him. “I also knew (and since I was immersed in the study of quantum theory, this part of his work interested me most keenly) that Einstein had produced a bold hypothesis on the subject of light. […] [In] every aspect of my studies I encountered with growing admiration the work of this lofty thinker” (\cite{LdB-1976}; p. 14).

After the Great War, Louis started to work with his brother Maurice. The elder brother recollected: “In the laboratory of rue Lord Byron, […] he found - with the work I was carrying out on the photoelectric effect of X-rays, a quantum phenomenon so curiously related both to waves and particles, with the company of young physicists focused on the corpuscular aspects of Roentgen’s rays - a fertile breeding ground for his ideas” \footnote{The usually cited address of Maurice's laboratory is 27 Rue Chateaubriand. Duke Philippe Maurice de Broglie confirmed that the two streets offered different accesses to the laboratory.}  (\cite{MdB-Penseur}, p. 426).

Louis’s work with Maurice and the investigations his brother prepared for Solvay III (1921) laid the groundwork for the wave-particle dualism for matter. Louis stressed the importance of the experimental work as follows (\cite{LdB-1931}; p.16): 
\begin{quote}
Nor should it be thought that, by focusing on X-rays, I was losing sight of my essential goal: studying quanta. Indeed, the whole theory of high-frequency radiation is imbued with the idea of the quantum. Like light, these radiations behave as if they were made up of corpuscles of energy $h\nu$, and the high value of the frequency means that the quantum $h\nu$ has a relatively large value, making it easier to demonstrate.

On the other hand, X-rays originate in the deep regions of the atom, the architecture of which is essentially determined by the quantum of action, as we learned from Bohr’s theory. Observation of the X-ray lines emitted by a given category of atoms provides information on the internal structure of these atoms and the movement of the electrons within them -- movements subject to quantum theory. Therefore, the study of X-ray spectra is closely linked to the theory of the atom and the quantum mechanics of electron motion in the atom. So it is easy to understand the interest it presented to me.
\end{quote}

\subsection{Preparing Solvay III: Atoms and Electrons (1921)}
Following his brother’s interests, Louis started investigating the interaction between X-rays and matter at the beginning of 1920. On March 31, the International Scientific Committee of the Solvay Institutes decided on the topic and the rapporteurs of the third Physics Council\footnote{For a detailed account, see the \textit{Procès Verbaux de la Commission Scientifique}, document L8/04, p. 105. PSL-ESPCI, \url{https://bibnum.explore.psl.eu/s/psl/ark:/18469/1z52k}.}. The next Council meeting was scheduled for April 1921. Madame Curie and Sir Rutherford were invited to propose the program while the Committee discussed the rapporteurs. One of the topics was \textit{Structure of the atom, charge, and constitution of the nucleus, isotopes.} Rutherford and Maurice were charged with preparing two distinct reports. Rutherford would indeed report on the structure of the atom at Solvay III while Maurice focused on the photoelectric effect for high frequencies, the absorption and emission of light and X-rays, and Einstein’s relation for light quanta: $ \mathcal{E} = h\nu $. 

After the Solvay Committee’s decision, Alexandre Dauviller joined Maurice’s laboratory. According to Wheaton, “Dauvillier’s research had already moved from studies of X-ray absorption to an examination of the kinetic energies of the secondary electrons released by X-rays” (\cite{Wheaton-book}, p. 265). However, the three researchers’ first published papers were on the absorption of X-rays. Maurice and Dauvillier focused on experimental facts, while Louis published his theoretical investigations elaborating Einstein’s probabilistic description of energy exchanges between matter and radiation, which he introduced in 1917. 

The investigations of this period focused on testing Bohr’s atomic model and explaining its tension with experimental measurements. However, at the beginning of 1921, the group’s investigations shifted to the so-called corpuscular spectra. Maurice published “A key finding” on January 31 (\cite{MdB-CR-172}, p. 275; January 31). Let us see how and why Maurice turned his research in this direction. He started his paper by summarizing the phenomena he was interested in (\cite{MdB-CR-172}, p. 274): 
\begin{quote}
    It is well known that bodies irradiated with X-rays will emit two different radiations: characteristic X-rays, called fluorescence rays, and very fast photo-electrics electrons. The new high-frequency spectroscopy investigates the first phenomenon; to date, the second has not been studied in depth; however, it offers similar interesting features. 
\end{quote}
As stressed by Wheaton, Maurice knew “the suggestive early work by Robinson and Rawlinson, and the incomplete experiments by Hu” (\cite{Wheaton-book}, p. 266), and he explicitly declared wanting to improve Hu’s results. In the already quoted paper, Maurice emphasized that “photoelectric electrons are characterized by their velocity” and that it was possible to investigate the velocity spectrum of the expelled particles using a magnetic field. Working with metals, Maurice analyzed Hu’s suggestion to connect their (relativistic) kinetic energy with Einstein’s formula $h\nu$. His preliminary work convinced him that he had found a corpuscular analog to specific X-rays\footnote{Maurice referred to K rays, which correspond to the X-rays emitted by stimulating the energy levels in the closest shell to the nucleus, the so-called \textit{K shell}.}. This correspondence impressed Maurice, who remarked: “This point completes the parallelism between the two emissions.” This parallelism was the first step leading to wave-particle dualism, and it resonated in his Solvay III report.

By improving his experimental apparatus, Maurice established that the energy conservation equation relating Einstein’s formula -- $\mathcal{E}=h\nu $ -- and the photoelectron kinetic energy is valid for almost-free electrons, the asymptotic region of X-ray spectra. The formula should be modified for bounded electrons to include the work to extract the bounded electron. He improved the experimental work and wrote a joint paper with Louis to give a theoretical basis to his experimental findings. On March 29, 1921, Maurice published the final proof of the connection between the corpuscular spectra, the work to extract them, and the atomic levels through the work needed to extract the electron from the atom. The day after, the Solvay Institutes received his report for Solvay III, to which Louis also contributed. Hence, it is unsurprising that Louis asked to be admitted as an auditor. As recollected by Maurice (\cite{MdB-1951}, p. 43):
\begin{quote}
    It was customary in Brussels to allow a few people from outside the Council -- but versed in physics -- to attend the sessions. A young man, almost unknown then, requested this authorization, which was refused. He keenly felt disappointed in this failure and vowed to return to the next Council, but this time through the front door. Louis de Broglie was to hold one of the first places in the Brussels meetings a few years later by exposing the new quantum dynamics that would restore the pace of theoretical conceptions rejuvenated by his work. 
\end{quote}
Lindemann, who had submitted a similar request, was also denied permission to attend the Council meeting.

\subsection{Discussions at Solvay III (1921)}
As recollected by Louis, the general result of his investigations with Maurice was that “the predictions of the corpuscular theory of radiation are exact in the domain of the X-photoelectric effect.” (\cite{LdB-1931}, p. 19) Hence, their work broadened the domain of the concept of light quanta and pointed towards its universality. The light quantum hypothesis was not universally accepted in the early 1920s before the discovery of the Compton effect. Their work for Solvay III on X-rays in Maurice’s well-equipped laboratory in the family’s mansion in Paris was also a starting step for Compton’s research.

Solvay III was devoted to \textit{Electrons, Atoms, and Radiation} and took place in Brussels from April 1 to 6, 1921. It welcomed W. L. Bragg and the American Nobel laureate Albert A. Michelson. Einstein was prevented from attending the meeting because of a trip to the United States. He nevertheless took part in the Council through an intermediary. His contribution was presented by Wander de Haas (Lorentz’s son-in-law). Bohr declined the invitation, being exhausted by the founding of the Copenhagen Institute of Theoretical Physics, which opened its doors on March 3, 1921. However, like Einstein, he participated in the Council through an intermediary. Ehrenfest presented his report. Being the first Council convened after WWI, Germans were not invited to Solvay III.

The debates focused on two fundamental issues. One was the Rutherford-Bohr model of the atom. The other was the dual wave-particle nature of light. In his report on \textit{The Structure of the Atom}, Rutherford recalled that it was only in 1914 that Charles G. Darwin showed that Hans W. Geiger and Ernest Marsden’s results on the scattering of alpha particles agreed with the customary law of electric force, which was derived by varying the velocity of incident particles.

Concerning radiation, there was a general feeling that Planck’s quantum of action, which had been shown to play a fundamental role in the constitution of atoms, should also be responsible for the complex nature of light and X-rays. This view, first introduced in 1905 with Einstein’s light quantum hypothesis in 1905 and reinforced in 1909 with his \textit{fluctuation formula}, was defended by Owen W. Richardson and Bragg. However, Maurice put forward the strongest arguments in favor of a particle-like nature of light.

Reporting on “pin-shaped X-rays,” Maurice discussed a series of phenomena - photoelectric effects and inverse processes - in which quanta seemed to intervene as “individual entities.” The discussion provided a long-awaited answer to the following remark made by Brillouin at the end of Solvay I (\cite{proceedings-Solvay1}, p. 451):
\begin{quote}
    It is very unsatisfactory to be reduced to recognizing the presence of a discontinuity by means of apparently continuous phenomena, to introduce it at the basis of a theory, before drowning it out with the help of statistical considerations [...] If we could imagine an experiment that made us seize the discontinuity \textit{on the spot}, it would be much more decisive and instructive” [emphasis added].
\end{quote}

In his report, Maurice commented on the particularly striking inability of classical wave theory to account for an experimental fact: “An atom that receives the light of frequency $\nu$ emits a projectile, endowed with energy $h\nu$, long before the radiation has been able to provide it with elements of this energy through homogeneous spherical waves” (\cite{proceedings-Solvay3}, p. 80). As Kuhn has stressed in the Foreword of Wheaton's book, Maurice insisted that X-ray radiation “must be corpuscular, or if it is undulatory, its energy must be concentrated in points of the surface of the wave” (\cite{Wheaton-book}, p. x).

One of the significant effects of Maurice’s work was its impact on understanding the photoelectric effect -- after Millikan’s well-known results -- in the X-ray frequency range. As noted by Wheaton, Maurice modified the Manchester $\beta$-ray velocity spectrometer and employed it to analyze X-ray absorption using what he called “spectres corpusculaires”, namely \textit{corpuscular spectra}. Wheaton remarked: 
\begin{quote}
    [By] introducing corpuscular spectra as a means to analyze the atom, he proceeded to a general discussion of the photoelectric effect. Assuming the validity of the Bohr-Kossel model for X-ray emission, it was clear that all the energy quantum of the X-ray is transferred to individual electrons. Coupled with the evidence that each X-ray frequency gives rise to a particular set of electron velocities, it became virtually impossible for [Maurice] de Broglie to avoid the issue of the light quantum (\cite{Wheaton-book}, p. 258).
\end{quote}

In fact, with his presentation at Solvay III, Maurice killed two birds with one stone: by validating Einstein’s expression for the energy of a light quantum, he provided decisive support to Bohr’s revolutionary postulates. Last but not least, Maurice’s work directly impacted the research of his young brother Louis. As reported in Wheaton’s book, Louis told his elder brother: 
\begin{quote}
    We debated the most pressing and baffling questions of the time, particularly the interpretation of your experiments on the X-ray photo effect [...] The insistence with which you directed my attention to the importance and undeniable accuracy of radiation's dual particle/wave properties little by little redirected my thought (\cite{Wheaton-book}, p. 275).
\end{quote}
Hence, to understand Louis’ turning point, we follow again Maurice’s work on X-rays.

Maurice’s report strongly asserted the symmetry between absorption and emission of light quanta. Regarding the absorption and emission of $\beta$-rays, he emphasized that corpuscular spectra are “kinetic transposition belonging to periodic vibrations” (\cite{proceedings-Solvay3}, p. 86). This means that he could (experimentally) reconstruct the X-spectrum from the kinetic-energy spectrum of $\beta-$particles by means of energy conservation. For the almost-free electrons of an irradiated gas, the kinetic energy corresponds to the irradiating X-ray, $h\nu$. As a consequence of this correspondence, during the discussion of Maurice’s report, Rutherford put forward the idea of a possible transformation process of X-rays into $\beta$-rays. In our view, Rutherford's remark influenced Maurice's viewpoint, as we shall see in the following. In the same discussion, Millikan emphasized how Maurice’s ideas would prove the existence of free electrons in metals. (\cite{proceedings-Solvay3}, p. 122).

An isolated remark by Léon Brillouin, the son of Marcel Brillouin, and Maurice's answer in the same discussion could have inspired Louis' future work. Brillouin discussed a thermodynamical analogy to treat blackbody radiation; Maurice hinted at the idea that the quantum could go under a degradation process after a series of absorption/emission processes ( See \cite{proceedings-Solvay3}, p. 114). Louis would develop this viewpoint after Solvay III.

\section{The legacy of Solvay III}\label{legacy}
Maurice gave us a clear idea of his feelings in 1921 when Louis started his great work. Reporting on Solvay III in his book, Maurice wrote \cite{MdB-1951}: 
\begin{quote}
    This was one of those periods when the balancing act between theory and experiment, the usual source of progress in the physical sciences, favored experiment, and it is clear from rereading Lorentz’s paper on the theory of electrons and the discussion that followed it, how much facts then overtook theoretical ideas.
\end{quote}
After Solvay III, Louis continued to collaborate with Dauvillier, but he returned to the subject of corpuscular spectra and the thermodynamic analogy we mentioned in analyzing Solvay III.

\subsection{Maurice’s perspective}
Maurice and Louis published two connected notes: \cite{MdB-1921} and \cite{LdB-1921a}. In his note, Maurice again stressed the perfect match between X-rays and corpuscular spectra, which can be interpreted as a “kinetic transposition of the spectrum belonging to periodic vibrations” (\cite{MdB-1921}, p. 1158). As already noticed by Olivier Darrigol, Maurice believed that “both electrons and X-rays possessed wave and corpuscular characteristics,” an idea triggered by the “extremely detailed correspondence between corpuscular spectrum and X-ray spectra” (\cite{Darrigol-1993}, p. 321). As we clarified in the previous section, the correspondence emerged from Maurice's experimental work for Solvay III and the discussion of his report. Rutherford's suggestion of a possible transmutation between the two, X-ray and $\beta-$particles, could have triggered the idea that electrons forming the corpuscular spectra possessed undulatory properties and reinforced the conviction that X-rays could behave as particles, paving the way to Louis' decision to describe electrons as a source of an undulatory phenomenon. However, it must be stressed that the elaboration of the theoretical leap that led to the wave-particle duality for electrons surely originated from Louis. As he recollected in an interview with Fritz Kubli in 1968, Maurice never liked to delve into the theory (\cite{Annales}, p. 23): 
\begin{quote}
I wrote this book while I was still working with my brother. I started to work less with him because I specialized in wave mechanics. [...] we had the idea of writing a book together on the physics of X-rays. \textit{My brother was, above all, an experimenter. He didn’t like theory very much}; he hadn’t studied it much. So, we shared the work. It was understood that I would take the theoretical chapters and that he would take the experimental chapters. 
\end{quote}
Before analyzing the meaning and importance of the thermodynamic analogy that emerged at Solvay III, we discuss the developments of Maurice’s work on X-rays and the idea highlighted by Darrigol.

Between the end of 1921 and 1922, Maurice worked on a book collecting his work and view on X-rays. Darrigol noticed that in this book: “[Maurice] willingly talked about the << frequency of the electrons >> meaning the frequency which, once multiplied by h, gave the (relativistic) energy of the electron” (\cite{Darrigol-1993}, p. 321). By digging deeper into Maurice’s book\footnote{In the following quotations of Maurice’s book, we are using its English translation dated 1925 after having checked nothing changed in comparison to the French version (1922).}, we can better understand his point of view. In his book, Maurice wrote (\cite{MdB-book-X}, p. 17): 
\begin{quote}
    Something kinetic is to be found in undulatory radiations and Something periodic in the projections of corpuscles, and all this makes it every day more tempting to think that a single reality is presented to us, sometimes as a kinetic and sometimes as an undulatory manifestation.
\end{quote}
In a footnote, Maurice stressed that “the emission of a radiation of a period which is well defined [...] \textit{corresponds} to corpuscles of a particular velocity, at the instant of their collision with matter” [emphasis added] (\cite{MdB-book-X}, p. 17). Hence, the undulatory nature of corpuscular spectra was not yet considered as an \textit{intrinsic} characteristic of the electrons, as in Louis’ future work, but a consequence of the interaction with electromagnetic radiation.

Maurice again expressed his intuitive idea based on energy conservation: “A quantum has a double significance; it indicates both the frequency of the radiation and a certain quantity of energy carried by the frequency. [...] A quantum can also be assigned to a particle possessing kinetic energy.” (\cite{MdB-book-X}, p. 18) Maurice finally added: “These considerations can be used to connect the kinetic and vibratory representations of radiations.”

To emphasize the unity of matter and radiation, Maurice presented a table (\cite{MdB-book-X}, p. 19) where he explicitly assigned a frequency and a wavelength to electrons moving at a speed of 38 m/s. The frequency can be defined by dividing the (non-relativistic) kinetic energy by the Planck constant. At the same time, the wavelength is obtained by dividing the speed of light by the frequency obtained. Hence, Maurice’s undulatory characteristics of electrons differed from the wavelength and frequency Louis will define. They were not yet intended as an intrinsic character of the electrons but defined through a quantity. This energy remains constant when, in contact with matter, beta and electromagnetic radiations are transformed into each other. However, Maurice insisted on the fact that his considerations of the quantum introduce “a kind of bond between undulatory and corpuscular radiations” (\cite{MdB-book-X}, p. 21). This bond seems to indirectly suggest that the wave-particle duality for light could be extended to electrons, as Louis would do. In the same table (\cite{MdB-book-X}, p. 19), Maurice computed the velocity of light quanta using energy conservation, obtaining a value that is less than $c$ and corresponds to the speed of the expelled electrons.

In his concluding remarks, Maurice emphasized the ideas regarding electromagnetic energy that  Einstein and Thomson had expressed at Solvay I. He stressed: “The fundamental fact is that the velocities of the expelled electrons always depend on the frequency of the radiation and not on its intensity. If it is necessary to assume, as has been sometimes suggested (Sir Joseph Thomson), that the energy is localized at specific points of the wavefront, we are very near to a corpuscular theory of light.” (\cite{MdB-book-X}, p. 156)

Another consequence of the centrality of Einstein’s ideas in Maurice's book is the following. If the energy of the electromagnetic wave is not uniformly distributed on the surface of the wavefront and we combine this idea “with the existence of an inertia for the energy, it will be seen how we get back to the corpuscular theory of radiation.” (\cite{MdB-book-X}, p. 20) In this last statement, Maurice suggests that it is possible to assign a mass to light quanta through Einstein’s formula relating mass and energy. The inertia associated with any form of energy is another ingredient Louis would use in his synthesis of wave-particle duality.

\subsection{Louis’ path to the photon gas}
We now reconsider the idea of degradation of light quanta suggested by Maurice at Solvay III and the two connected communications published in the \textit{Comptes-rendus de l’Académie des Sciences}, one by Maurice \cite{MdB-1921} and one by Louis \cite{LdB-1921a}. Maurice emphasized the equivalence between X-ray and corpuscular spectra from the experimental point of view.

Using the example of Tungsten, he described the correspondence between its X-ray spectrum of the K-level and the corpuscular spectrum represented by the alpha-, $(\alpha 1, \alpha2)$, and the two beta-, $(\beta 1, \beta 2)$, ray emissions. From Maurice’s perspective, the original quantum of energy had been split into quanta of lower frequency. Indeed, every corpuscular emission can be converted again into an X-ray quantum. In a sort of chain reaction, an X-ray quantum can be converted into a corpuscular emission, which can be converted into an X-ray mission, and so on. In this process, the original quantum drops down and is progressively degraded. In his communication \cite{LdB-1921a}, Louis analyzed a thermodynamical analogy linking the frequency with the temperature and the quantum of action with the entropy. Without entering into the technical details, this analogy played an essential role because it was Louis’s first attempt to treat X-ray emissions using thermodynamic concepts. This communication was the first step in applying statistical mechanics to X-ray emission.

Louis's following step was to work on the Bragg-Pierce law. In Ref. \cite{LdB-1921b}, Louis formulated his first step for X-ray absorption. Louis evaluated a constant of proportionality with the help of Bohr’s correspondence principle. Louis wrote: 
\begin{quote}
    Let us consider an ensemble of N atoms at a certain temperature T. We focus on very small frequencies for the temperature $T$. It is in the spirit of Bohr’s correspondence principle to admit that, for the emission of such frequencies, every atom can be treated as a tri-dimensional oscillator (\cite{LdB-1921b}, p. 1457).
\end{quote}
He used the result to compare his theoretical prediction with experimental data. The excellent matching of the two results led him to conclude that his calculation was showing “the far-reaching impact of the correspondence principle when it is interpreted as a statistical principle” (\cite{LdB-1921b}, p. 1458).

The last step is the paper he finished in June 1921 and published in February 1922 \cite{LdB-1922a}. This paper bridges his investigations on X-rays’ absorption and radiation’s structure. Louis added the elements of statistical mechanics investigating a closed system maintained at a fixed temperature formed by atoms and radiation. He quoted a work published by Brillouin, which was also mentioned in the proceedings of Solvay III, and where he quoted Einstein’s report of Solvay I. This fact is essential for two reasons.

First, it testifies that Louis was informed by his brother about the discussions of Solvay III. Second, Brillouin proved that statistical mechanics could be applied to continuous and discontinuous systems. Louis focused on the continuous spectrum of X-rays and Wien’s law. He applied statistical mechanics to the emitting electrons inside the atoms by modeling them as a gas. Using the quantum hypothesis, he obtained the Bragg-Pierce law. We will come back to this fact in the following.

Around the end of 1921 and the beginning of 1922, Louis turned his attention to the structure of radiation in two connected papers. In \cite{LdB-1922b}, received in January 1922, Louis derived Wien’s law. In his subsequent communication \cite{LdB-1922c}, he investigated the consequences of considering aggregates of light quanta to include interference phenomena in his corpuscular approach.

Let us now focus on the first paper. As Darrigol noticed, Louis used purely corpuscular assumptions. He represented the radiation as a gas of photons. More precisely, he wrote: “This work aims to establish a certain number of results known from radiation theory by a reasoning based solely on thermodynamics, kinetic theory, and quantum theory, without any intervention from electromagnetism.” [emphasis added] (\cite{LdB-1922b}, p. 33). This statement suggests interpreting this paper as a natural continuation of his work based on the use of the thermodynamical analogy, with the addition of elements of statistical mechanics.

In his recollections, Louis wrote, 
\begin{quote}
    My studies of high-frequency radiation had convinced me of the need to accept, at least in a certain sense, the granular structure of radiation. And yet the phenomena of interference and diffraction of light are there to show us the necessity of the wave theory, and in the very field of high-frequency radiation, isn’t the diffraction of X-rays by crystals, which has been the basis of all the progress made in X-ray spectroscopy, clearly a wave phenomenon? Preoccupied with these questions, I undertook a little work that was the prelude to the development of Wave Mechanics (\cite{LdB-1931}; p.21).
\end{quote}
Louis was referring to the work done by his brother, who had proved the need to introduce the quantum hypothesis in the range of X-ray frequencies. Louis described radiation as a gas of light quanta and applied statistical mechanics to it. Where did this idea come from?

In the previous paragraph, we stressed that Maurice suggested assigning a mass to the light quanta using Einstein’s relation. Maurice started writing his book in 1921 and published it around the end of 1922. The most recent paper published by Louis and quoted by Maurice in his book was published in February 1922. It is the work about the Bragg-Pierce law, which belongs to the last part of the book, while Maurice’s suggestion of a massive light quantum appears at the beginning of the book. These elements suggest that the two brothers discussed this idea. However, we cannot say if it originated from Maurice or Louis. We can say that Louis improved Maurice’s original idea: Louis assigned a velocity to the light quantum. In contrast, Maurice assigned a velocity to the beta-particle emerging after X-ray absorption by describing light quanta scattered by the walls of a cube. 

At the beginning of his article, Louis states that the mass of the light quantum is obtained by dividing its energy $h\nu$ by the square of the speed of light, $c^2$, as suggested by his brother, to give proof of the light pressure radiation formula by modeling light quanta as \textit{massive} particles scattering the wall of a cube. He didn’t specify if he was referring to the kinetic or the total energy. However, in the first footnote (\cite{LdB-1922b}, p. 422), Louis considered the Special Relativistic formulas of the kinetic energy $W$ and momentum $G$ for particles of rest mass $m_0$ and velocity $v$. He introduced a limiting procedure where $m_0$ tends to zero, and $v$ tends to $c$ to justify his assumption for the energy and obtain Einstein’s formula for the momentum of a photon. He imposed that, in this limit, the variable mass $\displaystyle m=\frac{m_0}{\sqrt{1-\frac{v^2}{c^2}}}$ would stay finite. Hence, the kinetic energy equals the total energy, and he could impose $W=h\nu=mc^2$. A similar reasoning for the momentum gives $\displaystyle G=\frac{m_0\beta c}{\sqrt{1-\frac{v^2}{c^2}}} = mc$ because $\beta=\frac{v}{c}$ tends to $1$. The two formulas combined in the main text gave him the correct formula for the radiation of pressure.

It must be stressed that Louis' approach is not entirely clear. Louis used the terms \textit{infinitely close to zero} and \textit{infinitely close to c}; hence, he seemed to be considering a nonzero rest mass $m_0$ and $v<c$. This interpretation appears to be confirmed by the fact that, as also noticed by Darrigol (\cite{Darrigol-1993}, p. 337), Louis would calculate the velocity $v$ of the light quantum at the end of the paper\footnote{We should point out that there must be a typo no one noticed in Louis' paper. Indeed, the correct formula should contain the \textit{fifth} power of the speed of light instead of the third power.}. However, in an interview with Kuhn, Louis recollected that he was thinking precisely of a limiting procedure, bothered by the fact that when v tends to c, the energy goes to infinity (\cite{LdB-interview}; part II, p. 5).

Louis recollected that, at the time, he was considering the \textit{physical} meaning of this limit, which, from his perspective, was empirically justifiable. A massive light quantum moving at a speed less than the speed of light reproduces Maxwell’s relation between energy and momentum only approximately, but the difference would not be measurable. This comment confirms that Louis considered light quanta to be relativistic particles, implicitly opening the possibility of extending the dualistic nature of light to electrons. However, from a strictly mathematical perspective, Louis' limit mass $m$ cannot exist when we interpret the formula for the total energy as a function of the two variables $m_0$ and $v$. Despite these inconsistencies, it is clear that Louis elaborated on his brother’s ideas and that this limiting procedure, giving the light radiation pressure formula, inspired Louis to identify light quanta with massive particles.

Another element explaining Louis’s inspiration emerges again from the discussions at the Solvay meetings. During the debate on Maurice’s report at Solvay III, Léon Brillouin presented one of his theoretical calculations published in 1920. Brillouin investigated the X-ray continuum emissions and absorption but focused on the velocity distribution of the decelerated or accelerated electrons that produce the phenomenon. Maurice noticed that Wien had already introduced the thermodynamic analogy in the context of X-rays in 1911 (\cite{proceedings-Solvay1}, p. 114). We know that Louis quoted Brillouin in his previous work and that in his work, Brillouin quoted Einstein’s report at Solvay I.

Furthermore, in his last paper, Louis informed us that he discussed his work with Léon Brillouin (\cite{LdB-1922b}, p. 426). All these elements suggest that Louis was stimulated to reread the proceedings of Solvay I, where Marcel Brillouin, Léon’s father, was present. Hence, the following question arises: are there some hints in Einstein’s report and the subsequent discussion? 

To answer our question, we first point out that Lorentz made the following comment: “There have often been protests against the application of statistical methods to radiation; I don’t see why these methods can’t be applied to radiation (see Lorentz’s report).” (\cite{proceedings-Solvay1}, p. 119) Once Louis had verified that the relativistic massive light quanta reproduced the correct radiation pressure formula, applying statistical mechanics would be a direct consequence.

As already said in section \ref{Louis-Solvay1}, Louis was an ardent admirer of Einstein’s work. In the proceedings of Solvay I, which Louis thoroughly studied, Einstein suggested the following: 
\begin{quote}
    Let us imagine a box with perfectly reflecting or white walls of volume $V$ containing radiation of energy $E$ that is substantially monochromatic and of frequency $\nu$ (...). Suppose the radiation’s density is sufficiently weak to remain in the domain of validity of Wien’s law. In that case, the statistical distribution law is the same as if the radiation were composed of distinct point particles, each possessing the energy $h\nu$. [...] If we imagine the radiation composed of small energy elements equal to $h\nu$, we immediately obtain an explanation for the probability law of diluted radiation (\cite{proceedings-Solvay1}, p. 442). 
\end{quote}
Again, Einstein emphasized: “If we imagine radiation made up of small energy elements equal to $h\nu$, we immediately obtain an explanation for the probability law of diluted radiation” (\cite{proceedings-Solvay1}, p. 443). Quoted Einstein’s comments suggest the possibility of treating radiation as an ensemble of particles.

The last hidden contribution of Solvay I comes from the footnote of page 423. Having used a limiting formulation of relativistic dynamics to describe light quanta, Louis could emphasize that Liouville’s theorem is applicable. Darrigol has observed that this fact is contained in Langevin’s relativity lectures (\cite{Darrigol-1993}; footnote 69 p. 335). However, we have already stressed that the importance of Liouville’s theorem, its connection with the canonical theory, and the applicability of Gibbs’ methods was one of the central issues discussed in the proceedings of Solvay I that Louis had studied very carefully.

Another argument that could have resonated with Einstein's ideas in Louis's mind could be the suggestion made by Lindemann in the paper Louis quoted at the end of his paper. In a short note published in Philosophical Magazine, Lindemann focused on the significance of the chemical constant and its relation to the behavior of gases at low temperatures. He pointed at an analogy between gases and radiation through the role played by the chemical constant on one side and the Stefan constant on the other. At the end of his paper, Lindemann suggested as following: “One may conclude, therefore, since the chemical constant can be replaced by radiation pressure, that its physical significance may perhaps be sought rather in the interaction of radiation and matter than in the subdivision of Gibbs’s N-dimensional space into finite elements of equal probability.” (\cite{Lindemann-1920}, p. 25) We could say that Louis took up this challenge in his work.

\subsection{Einstein in Paris}
Einstein visited Paris in March 1922 when Louis was working on his second communication on a gas of light quanta. Einstein lectured at the Sorbonne and the Collège de France. For Louis, this was an emotional moment, as he recalled it in the following: “During that period of my life, I caught a glimpse of Einstein, though only from afar […] While attending his lectures […], I was struck by his charm and facial expressions, sometimes meditative and aloof, lively and playful.” (\cite{LdB-1976}, p. 14)

Einstein’s visit influenced Louis' work. In his communication to the \textit{Comptes-rendus}, Louis explicitly quoted Einstein’s report at Solvay I. He reconsidered his fluctuation formula to show how introducing aggregates of light quanta could explain the first term, usually originating from Maxwell’s theory. Darrigol stressed that despite the result obtained by Louis, “[He] did not believe that the future theory would work with corpuscular theory only.” (\cite{Darrigol-1993}, p. 335) Louis' attitude is not surprising for two reasons. Firstly, because of Einstein’s influence. As already emphasized, at Solvay I, Einstein underlined that his formula would imply a coexistence of corpuscular and undulatory concepts. Second, a purely corpuscular theory of light could not explain interference phenomena unless some periodic phenomenon could be associated with the light particle. This problem will be the next Louis addressed. The following two papers contain the seed of this point of view.

In the work published in the \textit{Journal de Physique}, Louis obtained Wien’s law except for a coefficient 2, representing the possible polarization states of the photon, which is another aspect connected to the undulatory character of light. The following remark made by Louis clarifies that he was already considering light quanta as a source of a periodic phenomenon capable of explaining the polarization of light. 
\begin{quote}
    A complete theory of light quanta would have to introduce it [the polarization of light] in a form such as the following: each light atom would be linked to an internal state of right or left circular polarization represented by an axial vector having the direction of the propagation velocity [emphasis added]  (\cite{LdB-1922b}, p. 426).
\end{quote}
Louis' comment indicates he was convinced of the coexistence between particle and wave aspects like Einstein. It also tells us that he was already trying to define a periodic phenomenon originating from light particles. As noticed by Darrigol, Louis underlined in his communication to the \textit{Comptes-rendus} the need for a compromise by introducing the notion of periodicity into the theory of light quanta to take into account interference phenomena (\cite{Darrigol-1993}; p. 335). For the polarization of light, Louis imagined attaching an axial vector along the direction of the propagation vector (\cite{LdB-1922b}, p. 426)

Before proceeding, it should be pointed out that the deficiency of de Louis' particle approach in considering the polarization phenomena was recognized by Léon Brillouin. In his conferences at Collège de France, Léon pointed out that Einstein’s idea of grains of energy would not explain the polarization of light. Léon’s observation was another reason Louis retained the view of a particle’s coexistence and an undulatory picture, as only this last one could explain the polarization phenomena. Léon quoted the proceedings of Solvay I and Solvay II as essential references to understand the debate around Planck’s original hypothesis. He also referred to the proceedings of Solvay III and the conferences at Collège de France by Maurice for a deeper understanding of  Bohr’s atomic theory and its experimental verification. 

Louis would also refer to the Solvay conference proceedings in his PhD thesis. According to Léon Brillouin, Louis often discussed his ideas with him during 1921 and 1922. In his March 29, 1962 interview, Brillouin explained \cite{Brillouin-interview}: “We [Louis and Léon] discussed these ideas a dozen times. Several times. Sometimes he [Louis] came to see some papers in our library, sometimes we talked together for five minutes, sometimes for a longer time.” 

Léon also pointed out that “the most important part of the work was finished in 1922.” During the discussions with Brillouin, Louis had certainly absorbed Brillouin’s idea that in a dispersive medium, the energy of electromagnetic waves travels with the group velocity $U$, which is the speed of propagation of the amplitude’s variation\footnote{A similar topic was discussed by Lorentz and Einstein. See \textit{Einstein Papers}, Vol. 12, DOC. 298.}  (\cite{Brillouin-book}, p. 11). Louis would later associate a light quantum with a group of waves and identify the photon’s velocity with the group velocity U. In his interview, Brillouin insisted again on the early stage of Louis' work: 
\begin{quote}
    The crucial thing was when he discovered the relation between momentum and wavelength [...] Everything started from there [the papers of 1922]. That was the first mark. And the rest had to follow it. Each step took a long time of thinking and wondering before he decided how to adjust to a new term. But I am sure — we agree on that — the p relation, the lambda relation, was ‘22. And then, I remember, there was some time before his thesis could be printed  \cite{Brillouin-interview}. 
\end{quote}
It must be emphasized that de Broglie’s wavelength appeared after introducing the phase waves, which traveled at speed $V$, obeying the relation $UV=c^2$. Hence, Brillouin’s recollection seems partially incorrect. What can be inferred from Louis' Ref. \cite{LdB-1922b} is the relation $\displaystyle \lambda\approx\frac{h}{G}$, where $G$ is the momentum of a massive particle traveling at speed $U\approx c$ and for which we can write $\lambda\nu\approx U$. As already said, the idea of a massive photon triggered the idea to extend the relation between wavelength and momentum to other massive particles.

Marcel Brillouin, Léon’s father, lectured at Collège de France in the same period. Marcel had recently published the last of a series of papers where he elaborated on a proposal where the electrons would emit a wave while revolving around the nucleus. Louis did not mention Marcel’s work at this stage, but from the interviews with Louis and Léon -- 1962 and 1963 -- it emerges that Louis knew Marcel Brillouin’s proposal in 1922. However, Louis was not inspired by it. In his interview with Kubli, Louis admitted that maybe he inserted that quotation to please Léon Brillouin and that there is only a vague analogy between his and Marcel's old proposal \cite{Annales}.

Louis reverted to Einstein’s fluctuation formula in the communication dated November 1922 at the \textit{Comptes-rendus}. Louis suggested considering “Maxwell’s theory as a continuous approximation (limit) of the discontinuous structure of radiation” (\cite{LdB-1922c}, p. 812). As Darrigol noticed in this paper, Louis obtained Planck’s distribution law by summing the contribution of “light molecules” made of many quanta. For this reason, Louis imagined that the simultaneous presence of many quanta would be at the origin of the interference phenomena, but this statement was only a conjecture (\cite{LdB-1922c}, p. 66). 

On one side, the imperfect particle approach served as a basis for introducing periodic phenomena in the particle picture to describe light. Conversely, the suggestion to make the same extension for massive particles was implicit because the formula of relativistic dynamics involved a nonzero rest mass.

\subsection{The impact of De Broglie brothers on Compton’s work}
The study of secondary gamma and X-rays was one of the topics addressed by Arthur Compton in his report for the National Research Council. Compton’s account, published in October 1922, contains the first evidence of what we call nowadays the Compton effect. The research aimed to understand the softening of secondary gamma and X-rays. As pointed out by Compton, previous studies had shown that “this radiation was less penetrating than the primary beam, and that this difference in quality was greater the harder the primary rays employed.” (\cite{Compton-1922}, p. 15) For this reason, Compton investigated the frequency of secondary X-rays: spectroscopic measurement showed a distinct change in wavelength for the secondary rays detected at a right angle, while absorption measurement seemed to suggest that this secondary radiation was composed of two parts, “one having the same wavelength as the primary beam, and the other a slightly longer wavelength” (\cite{Compton-1922}, p. 16).

Compton’s reference to the quantum nature of the frequency shift was implicitly based on the results of Solvay III. Indeed, Compton started his discussion of this “Doppler effect in secondary X-rays” with the following statement: “Based on the quantum theory, a different hypothesis may be formed. Let us suppose that each electron, when it scatters X-rays, \textit{receives a whole quantum of energy} and reradiates the whole quantum in a definite direction” [emphasis added] (\cite{Compton-1922}, 18). As we have stressed above, the transmission of an entire quantum during the interaction between radiation and electrons was one of the central topics of Maurice’s report at Solvay III. Compton did not quote it, but he insisted on its result: the interaction between radiation and electrons must involve the absorption of an entire quantum (\cite{Compton-1922}; p. 26).

According to Compton, a consistent theory of X-rays was still lacking at the time. One of the problems pointed out by the American scientist was the explanation of X-ray absorption. In his report, Compton reviewed two possible explanations: a pure electromagnetic approach and Louis’ theory. The former relied upon Maxwell’s equation and gave the correct formula “without introducing the quantum concept” (\cite{Compton-1922}, p. 43). As we have already commented, Louis’s theory was instead based on the quanta hypothesis. However, this part of Compton’s report tells us that the two scientists were in contact. Indeed, Compton stated that he discussed this approach with Louis in exchanging letters (\cite{Compton-1922}, p. 44). Compton maintained a balanced approach in this report by presenting both perspectives. However, in his subsequent paper on \textit{Physical Review} where he describes the shift of wavelength as a scattering process, Compton quoted Louis’ explanation only (\cite{Compton-1923}, p. 496) and used his new experimental evidence with X-ray absorption to confirm the change in frequency. This fact testifies to Compton’s conviction of the quantum foundation of this effect.

In his later recollections, Compton remembered how difficult it was to accept the change in wavelength at the time. “The results, accurately confirming the theoretical predictions, immediately became a subject of the most lively scientific controversy I have ever known. I reported the results [...] before the American Physical Society in April 1923.” (\cite{Compton-1961}, p. 818). In this context, Maurice played again an important role. In his recollections, Compton stressed: “Having frequently repeated the experiments, I entered the debate with confidence but was nevertheless pleased to find the support  from P. A. Ross of Stanford and M. de Broglie of Paris, who had obtained photographic spectra showing results similar to mine.” (\cite{Compton-1961}, p. 818)

Einstein's following statement testifies to the importance of Compton’s result\footnote{See \textit{Einstein papers}, vol 14, DOC. 236.}: “The positive outcome of Compton’s experiment proves that radiation acts as if it were composed of discrete projectiles of energy, not just as regards energy transfer, but also as regards the collision effects.”

\subsection{The stroke of a genius}
1923 was that of Louis’ stroke of a genius, as Darrigol called it. In 1931, Louis recollected his path as follows. 
\begin{quote}
    The study of blackbody radiation strengthened my conviction that, to arrive at a more comprehensive theory of light and radiation, it was necessary to seek to unite the idea of corpuscles with that of waves. Reflecting on this point, I suddenly thought that a similar unification should also be realized in a theory of matter. [...] In the corpuscular conception of radiation, the fundamental formula is that of Einstein’s light quantum. This relationship between the energy of the radiation corpuscle and the frequency of the corresponding wave establishes a kind of bridge between the conception of waves and that of corpuscles; it creates a kind of correspondence between the two images. Hence, the constant $h$ establishes a link between the wave and the corpuscle. What do we see when we consider the theory of matter, particularly the theory of atomic systems? We see elementary particles of matter, particularly electrons, describing quantized motions whose specification involves the constant $h$ and integers. The appearance of integers immediately brings to mind interference and resonance phenomena, i.e., wave phenomena. Doesn’t this suggest that, in the theory of matter, waves should be added to corpuscles, the correspondence between one and the other being defined by formulas in which Planck’s constant plays an essential role? If we establish such a correspondence between waves and corpuscles in the case of matter, it might be identical to the one we must admit in the case of light. We would then have achieved a wonderful result: we would have created a general doctrine that establishes a correlation between waves and corpuscles in the realm of light as in that of matter. For light, this correspondence requires the coexistence of photons and light waves, the energy of a photon being related to the frequency of the wave by Einstein’s relation. Applied to matter, the same correspondence requires that every corpuscle of matter, every electron, for example, be associated with a wave, which accompanies and controls its motion. From that moment on, I thought it would be possible to obtain interference and diffraction phenomena with electrons (\cite{LdB-1931}, p. 23).
\end{quote}

According to Louis, the core of his innovative approach is contained in the three communications that appeared in the \textit{Comptes-rendus} \cite{LdB-1923a} \cite{LdB-1923b} \cite{LdB-1923c}. Again, as Louis recollected, we should emphasize that his starting point was Maurice’s suggestion to consider Einstein’s formula for mass and radiation $h\nu=m_0c^2$. Using the exact words in his brother’s book, Louis thought of this equivalence in the rest frame of a photon, which was a massive particle in his mind. From now on, Louis would no longer refer to the particle’s kinetic energy but its total energy. In this context, the constant $c$ is not the speed of light quanta but just “the speed limit of the theory of relativity.” (\cite{LdB-1923a}, p. 507)   

As Darrigol has observed, the crucial point in introducing the identity $h\nu=m_0c^2$ and the idea of an internal periodic phenomenon represented by an internal clock was that it produced a new theoretical difficulty. Louis introduced his law of accordance (or harmony) of phases to find a description consistent with relativistic formulas. According to Georges Lochack, during the celebrations of his eighties birthday at the Academy of Sciences, Louis said: “A man never has more than one great idea in his life. If I ever had such an idea, it is certainly the law of phase harmony.” (\cite{Lochak}, p. 4) Lochack underlined that the principal idea of Louis’ reasoning was the clock: “[This] is why [Louis] de Broglie always considered that the first fundamental idea of wave mechanics was [...] in the work of Newton [...] in which an internal frequency was already introduced in the particles of light” (\cite{Lochak}, p. 6).

We will not go into technical details that have been described elsewhere. Furthermore, we will not discuss the role of the Maupertuis and Fermat principle. We point out just some details. First, Louis obtained the relativistic version of Bohr’s orbits (\cite{LdB-1923a}, p. 507). Then, he identified the particle’s velocity with the group velocity and explained interference and diffraction (\cite{LdB-1923b}, p. 548).

In his discussion with Léon Brillouin, Louis should have learned that the energy of a light signal, i.e., a group of waves whose frequencies are nearly equal, travels with the group velocity (\cite{Brillouin-book} p. 12). On the contrary, the phase waves guiding the particle traveled at a speed greater than $c$. In Louis’ view, the energy must be localized, as Einstein suggested. The diffraction patterns are obtained by curving the particle’s trajectories at the cost of abandoning the conservation of momentum (\cite{LdB-1923b}, p. 549). Finally, he went back to the idea of photon gas. The presence of an accompanying wave made possible the introduction of a polarization vector (\cite{LdB-1923c}, 633).

Louis gathered all his results in his PhD thesis, which he defended on November 25, 1924. In his work, Louis quoted the three Solvay Councils in his historical introduction to discuss the energy fluctuations in black body radiation.

\section{Solvay IV (1924): Spreading the news}\label{Solvay4}
The International Scientific Committee members of the Solvay Institute met on April 18 and 19, 1923, to prepare the fourth Physics Council, scheduled for April 1924. The Committee then comprised the following members:  Lorentz, M. Curie, W. H. Bragg, E. Rutherford, M. Brillouin, H. Kamerlingh-Onnes, M. Knudsen, P. Langevin\footnote{Following the death of Righi on July 8, 1920, his seat in the ISC had been given to Langevin.}, and E. van Aubel. After lengthy discussions, the Committee decided to devote Solvay IV to \textit{the mechanism of metallic conductivity and a series of phenomena likely to shed light on that mechanism}. 
This choice was primarily due to Lorentz, who wished to select a subject that could be dealt with without German members\footnote{The banning of German scientists was connected to facts related to the World conflict. See Ref. \cite{Lambert-Berends}, p. 188.}. Einstein was on the list of possible invitees as a Swiss citizen. However, he begged Lorentz not to be invited\footnote{Just as his friend Max Born, Einstein was critical about the considerable reparation payments imposed by the Allies on the Weimar Republic. Einstein likely reacted, in part, to the occupation of the Ruhr by French and Belgian troops. See \cite{Born-Einstein-letters}, letters 29 and 45.}, arguing that he didn’t want to be the only German guest. In contrast, the Council welcomed George de Hevesy, a Hungarian researcher who worked in Copenhagen, and Erwin Schr\"odinger, an Austrian physicist attached to the University of Zurich. While the former was already well-known for his experimental contributions, the latter was highlighted for his theoretical skills. As already said, the Councils were intended as a place where theoretical and experimental physicists meet.

The idea of discussing the electrical conductivity of metals (an issue closely related to the \textit{electron theory}) came to Lorentz’s mind during his transatlantic tour from January to April 1922. This journey enabled him to meet first-rank American and Canadian physicists, such as Edwin Hall, the discoverer of the phenomenon that bears his name; Percy Bridgman, a specialist in high pressures; John MacLennan, a low-temperature expert from Toronto, and Compton, who was about to discover the \textit{Compton effect}.

Ernest Solvay died in May 1922, but his scientific project remained alive. Solvay IV took place from April 24 to 28, 1924, and the chosen topic was: \textit{Electric Conductivity in Metals}. Nine reports were on the Council’s agenda:   
\begin{itemize}
    \item[1] Electron theory's application to metals and their properties (Lorentz)
    \item[2] Conductivity phenomena in metals and their theoretical explanation      (Bridgman). 
    \item[3] A theory of conductivity in metals (Richardson). 
    \item[4] The internal structure of alloys (Rosenhain). 
    \item[5] Electrical resistance and metal expansion (Broniewski).
    \item[6] Electrical conductivity of crystals (Joffé). 
    \item[7] New experiments with superconductors (Keesom).  
    \item[8] Metal conductivity and magnetic field's transverse effects (Hall).
    \item[9] The propagation of radiation impulses (Joffé and Dobronravoff). 
\end{itemize}

Regarding the topics discussed, it should be noted that the difficulty of inviting German physicists, which prevented the Council from focusing on fundamental questions in quantum theory, did not stop the participants from exploring the limits of its application. In a text written on the occasion of Langevin’s death in December 1946, the Ukrainian physicist Abram Fedorovich Joffé, who is considered today as the “father” of Soviet Physics\footnote{It is more than probable that Joffé was invited at the suggestion of Langevin, a supporter of Soviet physics within the ISC. See Ref \cite{A-Langevin}.}, emphasized: “I remember a meeting with Paul Langevin in 1924 at the Solvay Congress, where the contradictions between classical and atomic physics appeared very clearly. Lorentz, the creator of the theory of electrons, could only express his regret at not having died five years earlier when the horizon of this theory seemed serene” (\cite{Joffe-1947}, 16).

The Stern-Gerlach experiment\footnote{Einstein described it at the time as “the most interesting advance” (\cite{Born-Einstein-letters}, letter 42).}, which showed the reality of angular momentum quantization in atomic systems, was briefly mentioned during Richardson's report (Einstein’s quantum relation was mentioned, as were the quantized Bohr orbits). As for Bohr’s second atomic model, it was addressed by Willem Keesom in his presentation of the Kamerlingh Onnes report\footnote{Keesom replaced Kamerlingh Onnes, who had fallen ill.}. In the ninth report, added at the very last moment, Joffé presented an experiment prepared with Dobronravoff that had not yet been published. The report again certified how the work of Maurice had definitively given proof of the localized character of the energy for light quanta, as Einstein suggested. In their report, the authors implicitly referred to the recently appeared proposal of Bohr, Kramers, and Slater, which was based on the idea that energy conservation should only be regarded as a statistical principle.

The participants discussed the different models explaining conduction in metals, crystals, and gases. Lorentz states, “It seems quite probable that conductivity must be due to a certain freedom of the peripheral electrons” (\cite{proceedings-Solvay4}; p. 123). However, some proposed models seemed to contradict Bohr’s atomic theory. In general, the concept of free electrons was difficult to reconcile with Bohr’s orbits. These tensions and all the unresolved problems regarding radiation could have stimulated the participants to discuss them in the corridors and during the breaks from the official discussion, as already happened\footnote{See Ref. \cite{Lambert-Berends}.}. In his reminiscences, Joffé wrote: “Because all participants lived together during the congresses and the committee's work and spoke to each other not only at the meetings but also in the evenings, they became closely acquainted with one another.” (\cite{Joffe-1962}, p. 51) At Solvay IV, Joffé, Debye, and Schr\"odinger spent much time together. Joffé wrote: “With Schr\"odinger and his wife, we had many conversations about political topics as early as 1924” (\cite{Joffe-1962}, p. 53). However, in the following, we argue that this could not be the only topic they discussed.

In 1924, Schr\"odinger had not developed \textit{wave mechanics}, a formalism based on the ideas put forward by Louis in his doctoral thesis, which was not published yet at the time of Solvay IV. However, Joffé probably informed Schr\"odinger and Debye of Louis’ work. According to Jammer: “Now, as we know from reliable sources, Langevin, despite any as yet recognized empirical verification, spoke about Louis’ work at the fourth Solvay Congress in April 1924” (\cite{Jammer-1989}, p. 248). One of Jammer’s “reliable sources” was Joffé, whose recollections are contained in two documents.

In the text written on the occasion of Langevin’s death \cite{Joffe-1947}, Joffé emphasized: “Langevin spoke about Louis’ work as an interesting attempt as synthesis, offering a possible way out of the impasse.” We know crucial discussions between participants in Solvay meetings often occurred outside the conference room. According to Joffé’s later recollections:
\begin{quote}
    In the spring of 1924, in Brussels, Langevin told me about Louis de Broglie’s ideas about electron waves. He had submitted the work for the doctoral award to de Broglie. Langevin was charmed by the insight and originality of de Broglie’s ideas, even if he did not believe they were real.” (\cite{Joffe-1962}, p. 64). 
\end{quote}
This comment, taken together with Joffé’s close friendship with Debye and Schr\"odinger, and with the Council members’ habits to discuss the latest developments in private, suggests that Schr\"odinger heard about de Louis’ thesis at Solvay IV.

There are indications that it was not until the fall of 1925 that Schr\"odinger paid serious attention to Louis’ idea (\cite{Joas}; p. 344, footnote 21). As Schr\"odinger testified, Einstein’s paper on Bose-Einstein statistics played a decisive role in this fact. However, Langevin might have played a role too. We report here the little story told by Edmond Bauer\footnote{Bauer did not give any temporal reference, but it seems plausible that the episode could have happened after the discussion of Louis’ thesis (November 1924).} \cite{Bauer-interview}. 
\begin{quote}
    I must tell you something else, which perhaps [Louis] de Broglie will not tell you about. [...] Victor Henri was an excellent friend of mine; he was also a professor in Z\"urich at the University. Weiss was at the Polytechnique. Victor Henri told me this story. He went to Paris from Zurich, and Langevin told him, “Oh, we [have] the very remarkable thesis of de Broglie, and I will give it to you.” Victor Henri returned to Zurich and began reading Louis de Broglie's thesis. [...] Schr\"odinger, after two weeks, gave it back to him and told him, That’s rubbish! [...] Victor Henri returned to Paris, saw Langevin, and told him what Schr\"odinger [had said]. And Langevin told him, I think Schr\"odinger is wrong; he must look at it again. Then Victor Henri came back to Zurich and told Schr\"odinger, You ought to read the thesis of de Broglie again; Langevin thinks this is an excellent work. Then Schr\"odinger reread it and began his work.
\end{quote}

Why did Langevin send a copy to Schr\"odinger? From what we said above, Langevin was trying to highlight Louis’ thesis, and he targeted Einsten, who was supposed to be at the Solvay Council, and the participants of Solvay IV. Jammer underlined he has been “unable to find additional independent evidence” of the story Bauer reminded \cite{Jammer-1989}. Indeed, there is no evidence for Schr\"odinger’s comment. However, when told about this episode, Louis emphasized that it is plausible, as suggested by one of the interviewers, that “[Schr\"odinger] refused to study your [de Broglie’s] thesis; [however] he had great admiration for Langevin. So Mr. Langevin thought your thesis was important, and he said to himself: well, if he insists, I’ll have to take a close look at it, and that’s that.” (\cite{LdB-interview}, p. 11)

\section{Solvay V: Electrons and Photons (1927)}\label{Solvay5}
In 1925, Louis published in the journal \textit{Philosophical Magazine} a short English version of the results contained in his PhD thesis. Werner Heisenberg, Pascual Jordan, and Max Born wrote a paper on Matrix Mechanics. Wolfgang Pauli formulated his exclusion principle. Around the end of 1925, Schr\"odinger discovered his wave equation, and in 1926, he published his papers on Wave Mechanics. 

The members of the ISC met in Brussels during the first days of April 1926 to fix the program of the fifth Physics Council. The Committee members agreed with Lorentz that the time was ripe for inviting \textit{moderate} German researchers. Germany’s decision to join the League of Nations facilitated the acceptance of these proposals by the Administrative Commission. However, the presence of Germans in Brussels was still a sensitive issue. It had to be approved by the King. On April 1, 1926, Lorentz was received in the audience by King Albert, who, according to Lorentz, declared: “A better understanding between peoples is necessary, and science should help to bring it about. Considering what the Germans have done in physics, it would be difficult to do [the Council] without them”\footnote{Letter to ISIP's Administrative Commission.}. The King's intervention settled the problem.

During the ISC meeting, it was decided that Solvay V would be devoted to \textit{Classical and Quantum Radiation Theories}. Seven reports were on the Council’s program. The Committee assigned Louis to present his approach reconciling the concept of light quanta with the undulatory nature of Maxwell waves. The last report was to be presented by Heisenberg or Schr\"odinger. This choice could sound strange because matrix mechanics, invented by Heisenberg, and wave mechanics, developed by Schr\"odinger on Louis’ ideas, were two different approaches. The reason for this alternative is that the participation of Heisenberg, a German fellow, was not sure, and Schr\"odinger had recently informed Lorentz he had found a connection between the approaches.

In 1927, Louis continued to elaborate on his ideas: “[To] really remain in agreement with classical ideas, it was necessary to succeed in incorporating the corpuscle into the wave in such a way as to obtain a clear and coherent image of the wave-particle dualism in the framework of space and time” (\cite{LdB-1947}, p. 183). Louis created his \textit{double-solution} approach\footnote{We will not enter into technical details. For an in-depth analysis of his work, see Ref. \cite{Bacciagaluppi}.}. However, Louis soon realized that his approach involved enormous mathematical difficulties apart from simple cases. Hence, as Louis recollected in Ref. \cite{LdB-1947}, he sought to establish himself more firmly on a sort of “fallback position that was less satisfactory in principle but perhaps easier to defend” (\cite{LdB-1947}, p. 185). Louis referred to his \textit{pilot-wave approach}\footnote{For a technical introduction to the pilot-wave approach, see Ref. \cite{Bacciagaluppi}.}. This evolution was connected with the Solvay meeting, as the following account testifies.
\begin{quote}
     Preparing a report for the fifth Solvay Physics Council, which was to meet in Brussels in October 1927, offered me this opportunity.

     The immense interest that the emergence of the new mechanics had aroused in scientific circles in all countries had led to the choice of the study of this doctrine in its various aspects as the theme of the planned meeting. The illustrious H.-A. Lorentz (who was to die at the beginning of the following year), responsible for organizing and presiding this Physics Council, wrote to me in June 1927 to ask me to send him a report on Wave Mechanics and its interpretation, report which was to be distributed to the members of the Council before its meeting and to provide one of the bases for the discussion. Convinced at that moment that my hypothesis of the double solution was too difficult to justify, I adopted a more mixed perspective in my report. I admitted, on the one hand, the existence of corpuscles imagined in a classical way as small point objects and, on the other, that of continuous associated waves transposed from physical optics into wave mechanics. The corpuscle placed in the wave was, therefore, found to coincide with an element of a fictitious fluid of probability [...]; my hypothesis was then that the corpuscle followed at all times the movement of the element of the fictitious fluid with which it coincided and the principle of interference resulted immediately. Thus, I placed the corpuscle within the wave in some authority, and I supposed that it was driven according to a well-defined law by the very propagation of the wave.

     [...] In my hypothesis, the wave “piloted” the corpuscle somehow, hence the name “pilot-wave theory” that I gave to this conception. This theory was undoubtedly much less complete than the one that I had hoped to be able to build with the hypothesis of the double solution since it no longer succeeded in incorporating the corpuscle into the wave and noted the wave-corpuscle dualism without attempting to delve deeper into its nature. However, in my eyes, it then had the advantage of retaining the traditional notion of a point corpuscle well localized in space and maintaining its movements’ rigorous determinism. This was the solution I gave an outline in my report to the Solvay Congress (\cite{LdB-1947}, p. 185).
\end{quote}
In our introduction, we stressed that the consequences of the idea of \textit{entangled states} were clarified after Solvay V.\footnote{The entanglement phenomenon was introduced by Schr\"odinger in the mid-1930s.} However, it is worth noticing that in 1927, Louis had already foreseen the existence of entangled states. The above quotation shows that he insisted that every single particle be assigned a distinct wave, even for a many-particle system. By criticizing Schr\"odinger's single wave in configuration space, Louis proposed that every particle maintain its individuality by attaching a guiding wave to every one of them. He was, therefore, aware that Schr\"odinger's single wave in configuration space could represent entangled states\footnote{Non-separable states, which are represented by wave functions that cannot be factorized, represent entangled states of two or more particles.}.

Lorentz described the specificity of the fifth Council in a Note published on October 23, 1927, by the Brussels newspaper \textit{Le Soir}.  
\begin{quote}
    The issues dealt with by the present Council are closely linked to those of sixteen years ago. Mr. Planck introduced the notion of discontinuities and abrupt transitions in the motion of atoms and electrons. In 1911, the question was to appreciate these discontinuities' role in various phenomena and examine the fundamental laws that govern them. The first exploration of new and vast terrain, the richness of which had been revealed during the years that have passed since then, already suggested the need to reform the foundations of mechanics to reveal the discontinuities in question (their magnitude is now referred to as “quanta”), and to characterize them, not as Something additional but as a fundamental and essential element. This year, the discussions will focus on the attempts made in recent years to develop what may be called “quantum mechanics,” in which Messrs. de Broglie, Heisenberg, Born, Schr\"odinger, Dirac, and others took part. The work of these physicists still constitutes a more or less disparate aspect because, despite the unity deep inside, significant divergences of opinion remain. As a result, the “clash of opinions” will not be lacking. In such circumstances, the Solvay method is particularly efficient and can clarify ideas and accelerate progress.
\end{quote}
Lorentz’s note brought Louis' name to the general public.

For Louis, Solvay V was crucial because he had the opportunity to meet Einstein. As Louis recollected: “The illustrious scientist [Einstein] [...] stressed the importance of the ideas underlying my thesis [...] The paper of Einstein’s drew the attention of scientists to my work [...] I have always felt that I owe him a great personal debt for the encouragement he gave me.” (\cite{LdB-1976}, p. 15) At Solvay V, Louis reported on wave mechanics. He recollected: “In addition to my intellectual pleasure at being asked to participate in the Congress, I was full of pleasure and curiosity at the prospect of meeting Albert Einstein and exchanging ideas with him. [...] During a fairly long talk, he made a profound impression on me and fully confirmed my expectations.” (\cite{LdB-1976}, p. 15)

During the fifth Solvay Physics Council, the debates were characterized by the confrontation between the classical and the quantum theories of radiation. Irving Langmuir provided us in a letter to his mother with his vivid account of the Council’s sessions and of the general feeling at the end of the debates\footnote{See Ref. \cite{Lambert-Berends}.}: 
\begin{quote}
   Strenuous meetings -- 8 hours per day in a lecture room and all evenings (often till one o’clock in the morning) in the hotel where we all stayed, discussing quantum theory. Wave mechanics is the universal mode of expression today. For the first time, quantum theory can be formulated in apparently complete form so that the discrepancy between the classical wave theory and quantum phenomena (even the photon theory) seems to be clearing up. 
\end{quote}

According to Louis: “Though the concepts introduced by wave mechanics were generally accepted at the time, the details of the physical interpretation of the wave-particle dualism remained very controversial. […] Schr\"odinger and myself tried to present a concrete and causal interpretation, by and large in agreement with the traditional concepts of physics […] Born, Bohr, Heisenberg, Pauli, and Dirac presented a novel, purely probabilistic interpretation based on Heisenberg’s recently discovered uncertainty relations.” (\cite{LdB-1976}, p. 15)

In Ref. \cite{LdB-1976}, Louis called the last group of physicists we mentioned “the indeterministic school.” They met his theory “with cold disapproval.” Lorentz, instead, “spoke out as a convinced partisan of determinism and concrete images of classical physics.” However, for Louis, the essential opinion was Eisntein’s. “Although he had never expressed the opinion that I found a definitive solution to the problem, he had always encouraged me in private conversations to persevere on my path.”

During Solvay V, a group of the participants went to Paris to celebrate the centenary of the death of Augustin Fresnel, a French scientist born in a little town in Normandy: Broglie. This town, where the two de Broglie brothers also grew up, still hosts a residence for de Broglie's family with the laboratory that Maurice used during his summer holidays. Louis and Einstein were among the participants who went to Paris for Fresnel's celebration. “On the return trip from Brussels to Paris […] I had a final conversation with Einstein on the arrival platform of the Gare du Nord. He told me that he had little confidence in the indeterminist interpretation and was worried about the exaggerated turn toward formalism that quantum physics was taking. Then, possibly going further than he might normally have liked to go, he told me that all physical theories, their mathematical expression apart, ought to lend themselves to so simple a description that even a child could understand them. And what could be less simple than the purely statistical interpretation offered by Wave Mechanics! Outside the station, he left me by saying: Carry on! You are on the right track!” [emphasis added] (\cite{LdB-1976}, p. 17) Einstein’s words should have resonated in Louis’ mind. They could explain why Louis never lost hope and, lately, tried again to develop his original ideas.

\section{Epilogue}
Lorentz died soon after the meeting in 1928. His presidency of the International Physics Committee coincided precisely with a process started in 1911 and culminated sixteen years later with what Heisenberg called \textit{the completion of quantum mechanics} \cite{Kuhn-1}: 
\begin{quote}
    In the autumn of 1927, the Solvay Conference took place in Brussels, and here, the new interpretation of quantum theory was exposed to the most ingenious criticism, particularly on the part of Einstein, and thereby received the crucial test. Since the Solvay Conference of 1927, the Copenhagen interpretation has been fairly generally accepted and has formed the basis of all practical applications of quantum theory.
\end{quote} 

As it is well known, Schr\"odinger and Einstein resisted the change for different reasons. As for Louis, some months after Solvay V, he changed his attitude and started teaching the new quantum mechanics. However, he never abandoned the idea of improving his approach. In 1947, Louis was aware of the weakness of his old attempts: “[As] I delved deeper into this theory of the pilot wave, I better perceived its fragility and difficulties” (\cite{LdB-1947}, p. 186). What happened? Why did he stop, notwithstanding Einstein's encouraging words? Between the end of 1927 and the beginning of 1928, one of his pupils, Léon Rosenfeld, had joined the indeterministic school in G\"ottingen. One reason explaining Louis' stop suggested in \cite{Rosenfeld} and based on Rosenfeld-de Broglie correspondence could be that Rosenfeld convinced Louis to embrace the new dogma. Louis' official position seems to confirm this standpoint. In 1947, Louis recollected as follows \cite{LdB-1947}: 
\begin{quote}
    [The] pilot wave theory [...] ultimately appeared irreconcilable with Born's principle. As the development of Wave Mechanics has shown more and more, [...] this development must necessarily lead to abandoning the classical representation of phenomena using precise images within the framework of space and time and, consequently, of the mechanical determinism linked to it. These thoughts began to impose themselves on me when, on the eve of the opening of the Solvay Council, I became aware of a memorandum published shortly before by Mr. Heisenberg where, guided by suggestions from Mr. Bohr, the young German scholar stated for the first time his famous relations of uncertainty and developed all the new ideas to which they are linked. But I still hesitated to adopt a viewpoint wholly opposed to all my habits and previous thoughts. The discussions on this subject at the Solvay Council were quite lively. [...]  \textit{Without letting myself be entirely convinced} by the arguments of MM. Bohr and Heisenberg, I began to appreciate all their importance and depth.

    At the beginning of 1928, I was convinced that, despite the enormous effort of intellectual recovery they entailed, it was necessary to adopt the conceptions of Bohr and Heisenberg. Invited to give lectures at the University of Hamburg in the spring of 1928, \textit{I gave my formal support for the new ideas} for the first time in public. [emphasis added]
\end{quote}
Louis’ doubts were well-founded; for example, reconciliation with Born's principle has been only recently established. However, Solvay V was not the end of the story, and new developments occurred in the 1950s. The emphasized parts of the above quotation testify to how Louis never fully embraced the new quantum mechanics. In another interview, Louis gave us another hint about what happened \cite{Annales}: “[At] that time (1927-1928), I was affected by circumstances that were \textit{completely unrelated to science}. My mother fell ill, and I had to look after her and so on, which prevented me from working. Then, in 1928, I was appointed professor at the Sorbonne.” All these events convinced Louis to focus on teaching, and since he felt that his approach was still incomplete, he disseminated the \textit{winning} approach.

Louis restarted working on the pilot-wave approach in the mid-1950s after Dadid Bohm revived his ideas. Novel discussions occurred, and new critiques led to the \textit{second quantum revolution}. However, this is another story. Despite all these future developments and the fact that some features of quantum mechanics have not yet been highlighted at Solvay V, Heisenberg's perception of the fifth Solvay Council as the closing event of a process we dubbed \textit{the first quantum revolution} was correct. This paper has highlighted the fundamental role played by Maurice and Louis de Broglie in connection with the stimulating actions of the International Solvay Institutes of Physics in the confirmation of the old quantum theory and the transition to quantum mechanics.

\begin{eref}


\bibitem{Hoffmann} D. Hoffmann, \textit{Erwin Schr\"odinger}, Leipzig, BSB B. G. Teubner Verlagsgesellschaft, 1984.

\bibitem{Kuhn-1} Interview of Werner Heisenberg by Thomas S. Kuhn and John Heilbron on 1962 November 30, Niels Bohr Library \& Archives, American Institute of Physics, College Park, MD USA

\bibitem{Lambert-Berends} F. J. Lambert and F. Berends, \textit{Einstein’s Witches’ Sabbath and the Early Solvay Councils. The Untold Story.}, Les Ulis, EDP Sciences, 2021.

\bibitem{Poincare} H. Poincaré, Sur la théorie des quanta, \textit{Comptes-rendus de l’Académie des Sciences}, \textbf{152}, 1103, (1911).

\bibitem{Wheaton-article} B. Wheaton, Atomic Waves in Private Practice. In: J. Evans and A. Thorndike, \textit{Quantum Mechanics at The Crossroads. New Perspectives from History, Philosophy and Physics}, Berlin-Heidelberg, Springer-Verlag, 2007.

\bibitem{LdB-1931} L. de Broglie, \textit{Notice sure les Travaux Scientifiques de M. Louis de Broglie}, Paris, Hermann et C.ie, 1931.

\bibitem{LdB-1976} L. de Broglie, My meeting with Einstein at the Solvay Conference of 1927. In: \textit{Einstein: A Centenary
Volume}, Harvard, Harvard University Press, 1976.

\bibitem{Minutes} M. de Broglie, \textit{Registre contenant des pièces manuscrites concernant les premiers Congrès Solvay, 2 volumes}, Archives de Académie des Sciences, Institut de France.

\bibitem{LdB-Penseur} L. de Broglie, Vue d’ensemble sur mes travaux scientifiques. In: \textit{Louis de Broglie. Physicien et
penseur}, Paris, Albin Michel, 1953.

\bibitem{MdB-Penseur} M. de Broglie, La jeunesse et less orientations intellectuelles de Louis de Broglie. In: \textit{Louis de Broglie. Physicien et penseur}, Paris, Albin Michel, 1953.

\bibitem{proceedings-Solvay1} P. Langevin and M. de Broglie, \textit{La théorie du rayonnement et les quanta. Rapports et discussions de la réunion tenue à Bruxelles du 30 octobre au 3 novembre 1911 sous les auspices de M. E. Solvay}, Paris, Gauthier-Villars, 1912.

\bibitem{MdB-1951} M. de Broglie, \textit{Les premiers congrès de Physique Solvay et l'orientation de la physique depuis 1911},
Sciences d'aujourd'hui, Paris, Albin Michel, 1951.

\bibitem{Trillat-1960} J. Trillat, Maurice de Broglie (1875-1960). \textit{Bulletin de la Société française de Minéralogie et de Cristallographie}, \textbf{83}, 238, (1960).

\bibitem{proceedings-Solvay3} International Solvay Instutute of Physics, \textit{Atomes et Electrons. Rapports et Discussions du Conseil de Physique tenu à Bruxelles du 1er au 6 avril 1921 sous les auspices de l’Institute International de Physique Solvay}, Paris, Gauthier-Villars, 1923.

\bibitem{Wheaton-book} B. Wheaton, \textit{The Tiger and the Shark. Empirical Roots of Wave-Particle Dualism}, Cambridge, Cambridge University Press, 1983.

\bibitem{MdB-CR-172} M. de Broglie, Sur les spectres corpusculaires des éléments, \textit{Comptes-rendus de l’Académie des Sciences}, \textbf{172}, 274, (1921).

\bibitem{MdB-1921} M. de Broglie, Sur les spectres corpusculaires et leur utilisation pour l'étude des spectres des
rayons X, \textit{Comptes-rendus de l’Académie des Sciences}, \textbf{173}, 1156, (1921).

\bibitem{LdB-1921a} L. de Broglie, Sur la dégradation du quantum dans les transformations successives
des radiations de haute fréquence, \textit{Comptes-rendus de l’Académie des Sciences}, \textbf{173}, 1160, (1921).

\bibitem{Darrigol-1993} O. Darrigol, Strangeness and Soundness in Louis de Broglie’s Early Work, \textit{Physis}, \textbf{30}, 303, (1993).

\bibitem{Annales} F. Kubli, An interview with Louis de Broglie, \textit{Annales de La Fondation Louis De Broglie}, \textbf{48-1}, 21, (2024).

\bibitem{MdB-book-X} M. de Broglie, \textit{Les rayons X}, Paris, Libr. Scient. Albert Blanchard, 1922, 20.

\bibitem{LdB-1921b} L. de Broglie, Sur la théorie de l'absorption des rayons X par la matière et le principe
de correspondence, \textit{Comptes-rendus de l’Académie des Sciences}, \textbf{173}, 1456, (1921).

\bibitem{LdB-1922a} L. de Broglie, Rayons X et équilibre thermodynamique, \textit{Journal de Physique}, \textbf{3},
33, (1922).

\bibitem{LdB-1922b} L. de Broglie, Rayonnement noir et quanta de lumière, \textit{Journal de Physique}, \textbf{3}, 422, (1922).

\bibitem{LdB-1922c} L. de Broglie, Sur les interférences et la théorie des quanta de lumière, \textit{Comptes-rendus de l’Académie des Sciences}, \textbf{175}, 811, (1922).

\bibitem{LdB-interview} Interview of Louis de Broglie by Thomas S. Kuhn, T. Kahan, and A. George on 1963 January 7 and 14, Niels Bohr Library \& Archives, American Institute of Physics, College Park, MD USA, \url{http://repository.aip.org/islandora/object/nbla:272502}.

\bibitem{Lindemann-1920} F. Lindemann, On the Significance of the Chemical Constant and its Relation to the Behaviour of Gases at Low Temperatures, \textit{Philosophycal Magazine}, \textbf{39}, 21, (1920).

\bibitem{Brillouin-interview} Interview of Léon Brillouin by Paul Peter Ewald with Thomas S. Kuhn, George Uhlenbeck, and Mrs. Ewald on 1962 March 29,
Niels Bohr Library \& Archives, American Institute of Physics, College Park, MD USA, \url{www.aip.org/history-programs/niels-bohr-library/oral-histories/4538-1}.

\bibitem{Brillouin-book} L. Brillouin, \textit{La theorie des quanta et L'atome de Bohr}, Paris, Society of the \textit{Journal de Physique}, 1922.

\bibitem{Compton-1922} A. H. Compton, Secondary Radiations produced by X-Rays, and some of their applications to physical problems, \textit{Bulletin of the National Research Council}, \textbf{20}, 190, (1922).

\bibitem{Compton-1923} A. H. Compton, A Quantum Theory of the Scattering of X—Rays by Light elements, \textit{Physical Review}, \textbf{21}, 483, (1923)

\bibitem{Compton-1961} A. H. Compton, The Scattering of X Rays as Particles, \textit{American Journal of Physics}, \textbf{29}, 817, (1961).

\bibitem{LdB-1923a} L. de Broglie, Ondes et Quanta, \textit{Comptes-rendus de l’Académie des Sciences}, \textbf{177}, 507, (1923).

\bibitem{LdB-1923b}  L. de Broglie, Quanta de lumière, diffraction et interférences, \textit{Comptes-rendus de l’Académie des Sciences}, \textbf{177}, 548, (1923).

\bibitem{LdB-1923c}  L. de Broglie, Les quanta, la théorie cinétique des gaz et le principe de Fermat, \textit{Comptes-rendus de l’Académie des Sciences}, \textbf{177}, 630, (1923).

\bibitem{Lochak} G. Lochak, De Broglie's initial conception of de Broglie waves. In: \textit{In The Wave-particle dualism: a tribute to Louis de Broglie on his 90th birthday}, Dordrecht, D. Reidel Publishing Company.

\bibitem{Born-Einstein-letters} A. Einstein and M. Born, \textit{Born-Einstein Letters, 1916-1955: Friendship, Politics and Physics in Uncertain Times.}, New York, Palgrave Macmillan, 2005.

\bibitem{A-Langevin} A. Langevin, Paul Langevin et les Congrès de Physique Solvay, \textit{La Pensée. Revue du Rationalisme Moderne}, 10, (1966).

\bibitem{Joffe-1947} A. Joffé, A la Mémoire d’un Maître et Ami, \textit{La Pensée- Revue du Rationalisme  Moderne}, \textbf{12}, 15, (1947).

\bibitem{proceedings-Solvay4} International Solvay Instutute of Physics, \textit{Condictibilité Électrique des Métaux et Problèmes Connexes. Rapports et Discussions du quatrième Conseil de Physique tenu à Bruxelles du 24 au 29 Avril 1924 sous les auspices de l’Institute International de Physique Solvay}, Paris, Gauthier-Villars, 1927.

\bibitem{Joffe-1962} A. F. Joffé, \textit{Begegnungen mit Physikern}, Leipzig , B. G. Teubner Verlagsgesellschaft, (1967).

\bibitem{Jammer-1989} M. Jammer, \textit{The Conceptual Development of Quantum Mechanics, The History of Modern Physics 1800-1950. Vol.12}, AIP, Tomash, 1989.

\bibitem{Joas} C. Joas and C. Lehner, The classical roots of wave mechanics: Schr\"odinger’s transformations of the optical-mechanical analogy, \textit{Studies in History and Philosophy of Science Part B: Studies in History and Philosophy of Modern Physics}, \textbf{40}, 338, (2009).

\bibitem{Bauer-interview} Interview of Edmond Bauer by Thomas S. Kuhn and Theo Kahan on 1963 January 8, Niels Bohr Library \& Archives, American Institute of Physics, College Park, MD USA, \url{www.aip.org/history-programs/niels-bohr-library/oral-histories/4498-1}.

\bibitem{LdB-1947} L. de Broglie, \textit{Physique et Microphysique}, Paris, Albin Micheal \'Editions, 1947.

\bibitem{Bacciagaluppi}
G. Bacciagaluppi and A. Valentini, \textit{Quantum Theory at the Crossroads: Reconsidering the 1927 Solvay Conference}, Cambridge, Cambridge University Press, 2009. 

\bibitem{Rosenfeld} A. S. Jacobsen, \textit{Léon Rosenfeld. Physics, Philosophy, and Politics in the Twentieth Century}, Singapore, World Scientific, 2012.

\end{eref}

\man{2 juill 2024}

\end{document}